\documentclass[prb,twocolumn,showpacs,preprintnumbers,amsmath,amssymb]{revtex4}
\usepackage{dcolumn}
\usepackage{bm}
\usepackage{amsmath,graphicx}
\usepackage{color}

\newcommand{\ep}{\varepsilon}
\newcommand{\pa}{\partial}

\begin{document}
\title{Manipulating the Tomonaga-Luttinger exponent by electric field modulation}
\author{Hiroyuki Shima}
\email{shima@eng.hokudai.ac.jp}
\affiliation{Department of Applied Physics, Graduate School of Engineering,
Hokkaido University, Sapporo 060-8628, Japan}
\affiliation{Department of Applied Mathematics 3, LaC\`aN, Universitat Polit\`echnica de Catalunya (UPC),
Barcelona 08034, Spain}

\author{Shota Ono}
\affiliation{Department of Applied Physics, Graduate School of Engineering,
Hokkaido University, Sapporo 060-8628, Japan}

\author{Hideo Yoshioka}
\affiliation{Department of Physics, Nara Women's University, Nara 630-8506, Japan}

\date{\today}

%
%

\begin{abstract}
We establish a theoretical framework for artificial control of
the power-law singularities in Tomonaga-Luttinger liquid states.
The exponent governing the power-law behaviors is found to increase
significantly with an increase in the amplitude of the periodic electric field modulation
applied externally to the system.
This field-induced shift in the exponent indicates the tunability of
the transport properties of quasi-one-dimensional electron systems.
\end{abstract}


\pacs{73.21.Hb, 71.10.Pm}



\maketitle

\section{Introduction}

Interacting electrons in one-dimensional (1D) metals constitute
a highly collective state of matter: the Tomonaga-Luttinger liquid (TLL) state.
\cite{Tomonaga,Luttinger,Voit_1995}
The collective nature of the TLL states is what distinguishes them from their higher-dimensional counterparts.
Interacting electrons in two or three dimensions form a Fermi liquid,
wherein the only effects of interaction are the modification of their effective mass
and the possibility of being scattered.
In one dimension, however, even the slightest correlation between electron motions has a dramatic effect,
leading to distinctive features that cannot be explained by the Fermi liquid theory.
To date, physical consequences of the TLL states have been experimentally
observed in various systems, including
carbon nanotubes,\cite{Bockrath_1999,Yao_1999,Bachtold_2001,Ishii_2003,Tombros_2006}
semiconducting quantum wires,\cite{Auslaender_2002,Tserkovnyak_2002_2003,Auslaender_2003,Steinberg_2008,Jompol_2009}
quasi-1D organic conductors,\cite{Schwartz_1998,Claessen_2002,Sing_2003}
quantum Hall edge states,\cite{Chang_1996}
and other materials having highly anisotropic conductivity.
\cite{CKim_1996,Segovia_1999,Slot_2004,Aleshin_2004,Hager_2005,Venkataraman_2006,BJKim_2006,Yuen_2009}
From the theoretical viewpoint, a more general TLL theory with a nonlinear dispersion
\cite{Imambekov_2009}
as well as a novel wave-packed dynamics through Y-shaped TLL junctions\cite{Tokuno_2008}
have been recently suggested.

A hallmark of TLL states is a pseudogap in the one-particle density of states $\nu(\ep)$
at the Fermi energy $\ep_F$.
Injection of an additional electron in the TLL ground state disrupts the pre-existing correlation,
thus requiring excitation of an infinite number of collective modes.
This results in a power-law singularity of the form $\nu(\ep) \propto |\ep-\ep_F|^{\alpha}$,
where $\alpha(>0)$ is called the TLL exponent.
The same power-law arises in the case of a differential tunneling
current \cite{Bockrath_1999}
$dI/dV \propto |V|^{\alpha}$ at high bias voltages $(eV \gg k_B T)$
and a temperature-dependent conductance
$G(T) \propto T^{\alpha}$ at low voltages $(eV \ll k_B T)$,
although $\alpha$ may change due to environment effects.\cite{Bubanja_2009}
These power-law behaviors are in strong contrast with the behavior of Fermi liquids;
in the case of Fermi liquids, $\nu(\ep)$ close to $\ep_F$ and $dI/dV$ become constant.

The TLL exponent $\alpha$ is nonuniversal; it is dependent on the interaction strength,\cite{Voit_1995}
the geometric shape of the system,\cite{Shima_2009}
and the position of tunneling.\cite{Fabrizio_1995}
In fact, different values of $\alpha$
were obtained in carbon nanotube experiments
when each electron tunneled into the end or bulk of the system.\cite{Bockrath_1999,Yao_1999,Postma_2000}
A further non-trivial shift in $\alpha$ was suggested in multiwalled nanotubes, 
where $\alpha$ varies in a continuous manner under the application of a high transverse magnetic field.\cite{Bellucci_2006}
Continuous variation in $\alpha$ was also found
in a nuclear magnetic resonance study of CuBr$_4$(C$_5$H$_{12}$N)$_2$
crystals;\cite{Klanjsek_2008}
in this case, an external magnetic field acted as the chemical potential.
Such field-induced variations in $\alpha$
can be exploited for achieving artificial control of transport properties in quasi-1D conductors,
which would play a fundamental role in the development of next-generation quantum devices.

In this paper, we propose a theoretical framework for TLL exponent manipulation
based on electric field modulation.
An analytical expression of the exponent $\alpha$
for quasi-1D conductors subjected to a stepwise periodic potential
is established in terms of the potential amplitude and period.
Under feasible physical conditions, $\alpha$ increases significantly with an increase in the potential amplitude; this indicates that it is possible to tune the quantum transport properties of quasi-1D systems by manipulating $\alpha$.
For a concise description,
we focus our attention on 1D spinless fermion systems,
considering that the effects of spin degree of freedom
requires no substantial revision of the present conclusion.
This issue will be revisited in Sec.~IV.

\section{One-particle states in 1D periodic systems}

\subsection{Eigenenergy analysis}

We consider a quasi-1D electron system having a thin cylindrical shape
with length $L$, which is subjected to a periodic external potential field.
The cylinder radius $d$ is so small $(d\ll L)$ that
all electrons
reside in the lowest subband, $\chi$, of the transverse motion.
Single-particle wavefunctions thus have the form
\begin{equation}
\Psi(z, \bm{r}_{\perp})
= \chi(\bm{r}_{\perp}) \psi(z),
\label{eq_psi}
\end{equation}
where the two-dimensional vector $\bm{r}_{\perp} = (x,y)$ spans the circular cross section.
The axial component $\psi(z)$ obeys the Schr\"odinger equation
with effective mass $m^*$,
\begin{equation}
-\frac{\hbar^2}{2m^*} \frac{d^2 \psi}{d z^2} +  u(z) \psi(z) = \ep \psi(z),
\label{eq_01}
\end{equation}
where $u$ is a stepwise periodic potential given by
\begin{equation}
u(z) = \left\{
\begin{array}{lc}
0, & (0\le z <a) \\
u_0>0, & (-b \le z<0)
\end{array}
\right.
\label{eq_02}
\end{equation}
and $u(z) = u(z+a+b)$.
The periodicity of $u(z)$ implies that
the eigenstates of Eq.~(\ref{eq_01}) are represented by the Bloch function
\begin{equation}
\psi(z) = \frac{1}{\sqrt{N}} \phi_k(z) e^{ikz}, \;\;\; \phi_k(z+a+b) = \phi_k(z),
\label{eq_002b}
\end{equation}
where $\int_0^L |\psi(z)|^2 dz = 1$ and $\int_0^{a+b} |\phi(z)|^2 dz = 1$ with $L=N(a+b)$.
Hence, all eigenstates are labeled by the index $k$.
It is noteworthy that the periodic potentials similar to the above
can be realized by introducing geometric curvature
(instead of field modulation) to quati-1D systems,
and such the curved systems may show non-trivial
quantum transport.\cite{Ono,Taira}

The dispersion relation of the system given by Eq.~(\ref{eq_01})
is (see Appendix A)
\begin{equation}
f(\ep) = \cos \left[ k (a+b)\right].
\label{eq_05}
\end{equation}
Here, the function $f(\ep)$ is given by
\begin{eqnarray}
f(\ep)
&=& \frac12\left( \frac{\eta_1}{\eta_0} - \frac{\eta_0}{\eta_1} \right)
\sin \left( \zeta_a \eta_0 \right)
\sinh \left( \zeta_b \eta_1 \right) \nonumber \\ [2mm]
& &+
\cos \left( \zeta_a \eta_0 \right)
\cosh \left( \zeta_b \eta_1 \right),
\label{eq_06x}
\end{eqnarray}
where $\eta_0 = \sqrt{\ep}$, $\eta_1 = \sqrt{u_0 - \ep}$,
$\zeta_a^2 = 2m^* a^2/\hbar^2$, and $\zeta_b^2 = 2m^* b^2/\hbar^2$.
According to Eq. (\ref{eq_05}), $f(\ep)$ must fall in the range $-1$ to $1$
for $\ep$ to be physically relevant;
{\it i.e.}, only $\ep$'s that satisfy the condition
$|f(\ep)|\le 1$ are allowed to be the eigenenergies of the system.
Note that Eq.~(\ref{eq_05})
reduces to the trivial relation $\ep = \hbar^2 k^2/(2m^*)$
when $u_0 = 0$, since Eq.~(\ref{eq_06x}) becomes
$f(\ep, u_0=0) = \cos[(\zeta_a+\zeta_b)\sqrt{\ep}]$.
Throughout this paper, we use units of the effective Bohr radius $a_{\rm B} = \hbar^2 \epsilon/(m^* e^2)$
for length and units of the effective Rydberg ${\rm Ry} = e^2/(\epsilon a_{\rm B})$ for energy,
where $\epsilon$ is the dielectric constant of the wire.
For GaAs-based quantum wires, for instance,
we have $a_{\rm B} \sim 10$ nm and ${\rm Ry} \sim 50$ meV.\cite{Adachi_1985}

\begin{figure}[ttt]
\includegraphics[width=8.5cm]{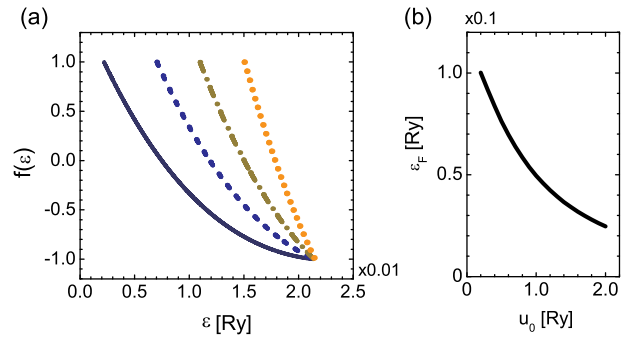}
\caption{(a) Variation of the curve $f(\ep)$ with an increase in the potential amplitude from
$u_0 = 0.2$ (solid) to $0.5$ (dashed), $1.0$ (dashed-dotted), and
$2.0$ (dotted) in units of
${\rm Ry} \equiv e^2/(\epsilon a_{\rm B})$ wherein $a_{\rm B} = \hbar^2 \epsilon/(m^* e^2)$.
The parameters $a=5.0$ and $b=1.0$ in units of $a_{\rm B}$ are fixed.
(b) Slow decay of $\ep_F$ with increasing $u_0$;
see text for the definition of $\ep_F$.
}
\label{fig_01}
\end{figure}

Figure \ref{fig_01}(a) shows the plot of $f(\ep)$ corresponding to the lowest energy band of the allowed $\ep$.
The potential amplitude $u_0$ ranges from 0.2 to 2.0,
and fixed parameters $a=5.0$ and $b=1.0$ are used.
With increasing $u_0$,
the band width shrinks monotonically and the ground-state energy $\ep_{\rm grd}$
({\it i.e.}, the specific $\ep$ that gives $f(\ep) = 1$ in the plot)
shifts to the right.
Thereafter, we set the Fermi energy $\ep_F$ (measured from $\ep_{\rm grd}$) such that
the electron density $n$ satisfies $n a_{\rm B} = 0.3$,
which gave $r_s \equiv (2 n a_{\rm B})^{-1} = 1.67$.
This value is much lower than the critical value of $r_s$, {\it viz.,} $r_s = 36$,
at which the Wigner crystal transition takes place\cite{Tanatar_1989};
it is also lower than $r_s \sim 2.2$, above which
the $4 k_F$ correlation in the charge density distribution
was suggested.\cite{Shulenburger_2008}
Since $n=(2/\pi) \int_0^{k_F} dk$, we can evaluate $\ep_F$ for a given $u_0$
from the relation $f(\ep_F) = \cos [\frac{\pi}{2}n(a+b)]$; see Eq.~(\ref{eq_05}).
Figure \ref{fig_01}(b) presents the $u_0$-dependencies of
$\ep_F$.
It decays with increasing $u_0$ but remains of the order of $10^{-2}$ Ry
for all $u_0$ under consideration.

\subsection{Field-induced shift in Fermi velocity}

The group velocity $v_g(\ep) \equiv \hbar^{-1} \pa \ep/\pa k$ of the system
is evaluated by differentiating both sides of Eq.~(\ref{eq_05}) with respect to $k$.
The result is
\begin{equation}
v_g(\ep) = - \frac{a+b}{\hbar}\frac{\sqrt{1-f(\ep)^2}}{\pa f/\pa \ep},
\label{eq_008}
\end{equation}
The explicit form of $\pa f/\pa \ep$ is given in Eq.~(\ref{eq_a012}) in Appendix A.
$v_g$ vanishes when $\ep$ satisfies $f(\ep) = \pm 1$,
{\it i.e.}, at the lower and upper band edges;
between the two edges, $v_g$ takes the maximum value.
The only exception occurs in the limit of $u_0 = 0$,
in which $v_g(\ep) = \sqrt{2\ep/m^*}$ increases monotonically with $\ep$.

\begin{figure}[ttt]
\includegraphics[width=8.0cm]{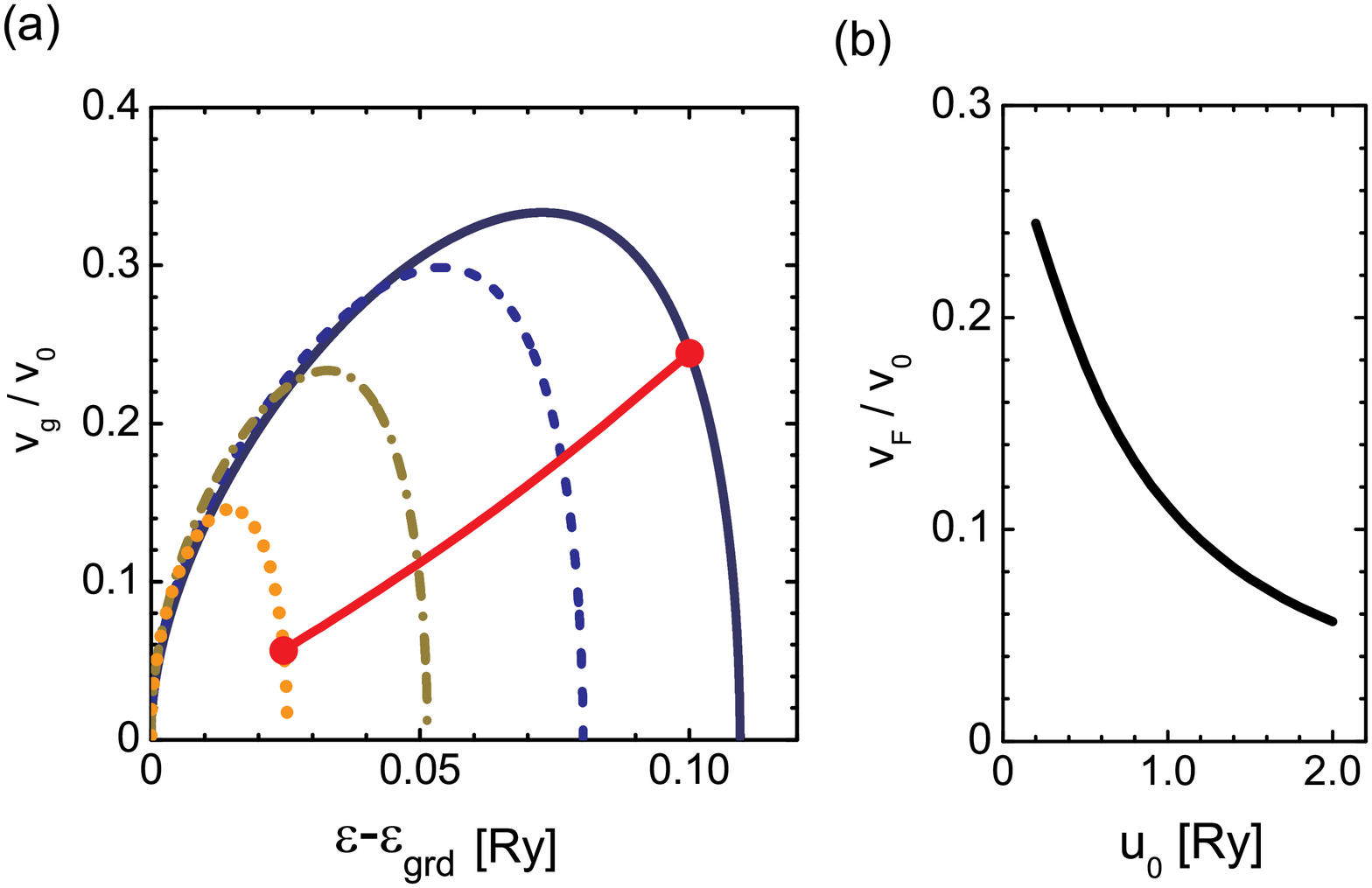}
\caption{(a) Profiles of the group velocity $v_g(\ep)$
in units of $v_0 \equiv {\rm Ry} a_{\rm B}/\hbar$.
Each curve corresponds to the value of $u_0$ in the same manner as Fig.~\ref{fig_01}.
A concave curve connecting the two end points
is the trajectory of the Fermi velocity
$v_F \equiv v_g(\ep=\ep_F)$
obtained through the $u_0$ variation.
(b) The Fermi velocity $v_F$ as a function of $u_0$.}
\label{fig_02}
\end{figure}

Figure \ref{fig_02}(a) shows the $\ep$-dependence of $v_g$
in units of $v_0 \equiv {\rm Ry} a_{\rm B}/\hbar$.
$u_0$ varies from 0.2 to 2.0 in the same manner as in Fig.~\ref{fig_01}.
With increasing $u_0$, the maximum value of $v_g$ decreases and the peak position
shifts to a lower $\ep$.
A concave curve connecting the two end points
is the trajectory of the Fermi velocity
$v_F \equiv v_g(\ep=\ep_F)$
obtained through the $u_0$ variation.
The value of $v_F$ decreases monotonically as $u_0$ increases,
and it then converges to the origin in the limit of $u_0 \to \infty$.
Figure \ref{fig_02}(b) shows the plot of $v_F$ vs. $u_0$.
Note that $v_F/v_0 \sim 0.1$ corresponds to $v_F \sim 10^5$ m/s
if we employ the material constants of GaAs.\cite{Adachi_1985}

\section{Bosonization of 1D periodic systems}

\subsection{TLL exponent in 1D periodic system}

Here we discuss the effect of field modulation on the TLL states.
An important indicator of TLL state realization is
a power-law singularity of the one-particle density of states $\nu(\ep)$
near $\ep_F$ represented by \cite{Voit_1995}
\begin{equation}
\nu(\ep)\propto |\ep - \ep_F|^{\alpha},
\quad \alpha = \frac12 \left( {\cal K}+ \frac{1}{{\cal K}} \right) -1.
\label{eq_025}
\end{equation}
Here, $\alpha$ is the TLL exponent,
and the parameter ${\cal K}$ is defined as (see Appendix B)
\begin{equation}
{\cal K} =
\left[ \frac{2 \pi \hbar v_F + (g_4 - g_2 + g_1)}{2\pi \hbar v_F + (g_4 + g_2 - g_1)} \right]^{1/2},
\label{eq_030}
\end{equation}
where
\begin{eqnarray}
g_4 &=& L \tilde{V} (k_F, k_F;\; q_z=0), \label{eq_g4} \\
g_2 &=& L \tilde{V} (k_F, -k_F;\; q_z= 0), \label{eq_g2} \\
g_1 &=& \frac{L}{2} \left[ V(k_F, -k_F; \; q_z = -2k_F) \right. \nonumber \\
& & \;\; \left. + V(-k_F, k_F; \; q_z = 2 k_F) \right] \label{eq_g1}
\end{eqnarray}
and
\begin{equation}
\tilde{V} (k_1, k_2, q_z) =
\left. \left\langle k_1+ q_z; \; k_2 - q_z \right. \right|
\hat{V}
\left. \left| k_2; \; k_1 \right. \right\rangle.
\label{eq_element}
\end{equation}
The right side of (\ref{eq_element}) is the matrix element of the Coulomb interaction between two-electron states
$\langle \bm{r}_j | k_j \rangle \equiv \Psi(\bm{r}_j)$
given by Eq.~(\ref{eq_psi}),
where $q_z$ describes the momentum transfer in the axial $(z)$ direction.
It should be noted that in formula (\ref{eq_030}),
we can prove that $g_4 = g_2$ as a consequence of the symmetric property
$\phi_k(z) = \phi_{-k}^*(z)$
of the function $\phi_k(z)$ introduced in Eq.~(\ref{eq_002b}).
See Eq.~(\ref{eq_002xx}) for the proof.

\subsection{Fourier representation of Coulomb interaction}

Computation of the TLL exponent $\alpha$ requires
evaluations of $\tilde{V} (k_1, k_2, q_z)$ at the specific values
of $k_1, k_2, q_z$ given above.
The explicit form of $\tilde{V} (k_1, k_2, q_z)$ is obtained by assuming the screened Coulomb potential
\begin{equation}
\langle \bm{r}_i, \bm{r}_j | \hat{V} | \bm{r}_j, \bm{r}_i \rangle
=
V(\bm{r}_{ij})
\; = \;
c_{\epsilon} \frac{e^{-\kappa |\bm{r}_{ij}|}}{|\bm{r}_{ij}|},
\label{eq_vr}
\end{equation}
where $c_{\epsilon} = {\rm e}^2/(4\pi \epsilon)$ and
$\kappa$ is the inverse of the screening length.
Equation (\ref{eq_vr}) has an alternative expression, given as
\begin{equation}
V(\bm{r}_{ij})
=
\frac{1}{(2\pi)^3} \int d\bm{q} \frac{4\pi c_{\epsilon}}{|\bm{q}|^2 + \kappa^2}
e^{i \bm{q}\cdot \bm{r}_{ij}}.
\label{eq_022b}
\end{equation}
From Eqs.~(\ref{eq_element}) and (\ref{eq_022b}), we obtain
(see also Eq.~(\ref{eq_appb12}))
\begin{eqnarray}
\!\!\!& &\tilde{V}(k_1, k_2, q_z)
=
\frac{\lambda}{L} \int \frac{dq_z'}{2\pi} \int \frac{d\bm{q}_{\perp}'}{(2\pi)^2}  \nonumber \\
\!\!\! & & \times
\frac{4\pi c_{\epsilon} |F(\bm{q}_{\perp}')|^2}{(q_z')^2 + |\bm{q}_{\perp}'|^2 + \kappa^2}
G(q_z, q_z', k_1) G_U(q_z, q_z', k_2),
\label{eq_056}
\end{eqnarray}
where $\lambda=a+b$, $\bm{q}_{\perp} =(q_x,q_y)$ is
the transverse component of the three-dimensional
wavevector $\bm{q}$, and
\begin{eqnarray}
\!\!\!\!\! F(\bm{q}_{\perp}')
\!\! &=& \!\!\!
\int \!\! d\bm{r}_{\perp} |\chi(\bm{r}_{\perp})|^2 e^{-i \bm{q}_{\perp}' \cdot \bm{r}_{\perp}}, \\
\!\!\!\!\! G(q_z, q_z', k_1)
\!\! &=& \!\!\!
\int \!\! dz_1 \phi_{k_1 + q_z}^*(z_1) \phi_{k_1}(z_1) e^{-i (q_z-q_z') z_1}, \\
\!\!\!\!\! G_U(q_z, q_z', k_2)
\!\! &=& \!\!\!
\int_U \!\! dz_2 \phi_{k_2 - q_z}^*(z_2) \phi_{k_2}(z_2) e^{i (q_z-q_z') z_2}.\label{eq_027}
\end{eqnarray}
The subscript $U$ in $G_U$ and $\int_U$ in Eq.~(\ref{eq_027}) indicates
integration within the unit cell domain $z_2\in [0,\lambda]$.
It is natural to assume that $\chi(\bm{r}_{\perp})$ has a Gaussian form such as\cite{Taira}
\begin{equation}
\chi(\bm{r}_{\perp}) = \left( \frac{2}{\pi d^2} \right)^{1/2}
\exp \left(- \frac{|\bm{r}_{\perp}|^2}{d^2} \right),
\label{eq_025b}
\end{equation}
which gives $F(\bm{q}_{\perp}') = \exp\left(-|\bm{q}_{\perp}'|^2 d^2/2\right)$.
As a result, the integral term with respect to $\bm{q}_{\perp}'$ in Eq.~(\ref{eq_056})
is rewritten as
\begin{eqnarray}
\!\!\!\!\! & &\int d\bm{q}_{\perp}'
\frac{ e^{-|\bm{q}_{\perp}'|^2 d^2/2}}{(q_z')^2 + |\bm{q}_{\perp}'|^2 + \kappa^2}
=
\int_0^{\infty} dt \frac{\pi e^{-t}}{t + \left( {q_z'}^2 + \kappa^2 \right) d^2} \nonumber \\
\!\!\!\!\! & &=
-\pi e^{\left( {q_z'}^2 + \kappa^2 \right) d^2 } \;
{\rm Ei} \left[- \left( {q_z'}^2 + \kappa^2 \right) d^2 \right],
\end{eqnarray}
where ${\rm Ei}(x)$ is the exponential-integral function
defined by ${\rm Ei}(x) = - \int_{-x}^{\infty} t^{-1} e^{-t} dt$.
Finally, we obtain the explicit form
\begin{eqnarray}
\tilde{V}(k_1, k_2, q_z)
&=&
- \frac{c_{\epsilon} \lambda}{4\pi L}
\int dq_z' G(q_z, q_z', k_1) G_U(q_z, q_z', k_2) \nonumber \\
&\times&
e^{\left( {q_z'}^2 + \kappa^2 \right) d^2 } \;
{\rm Ei} \left[- \left( {q_z'}^2 + \kappa^2 \right) d^2 \right],
\label{eq_002xx}
\end{eqnarray}
by which we can compute $g_4$, $g_2$, $g_1$ in Eqs.~(\ref{eq_g4})-(\ref{eq_g1})
for a given $k_F$.

\section{Results and discussions}

\begin{figure}[ttt]
\includegraphics[width=7.5cm]{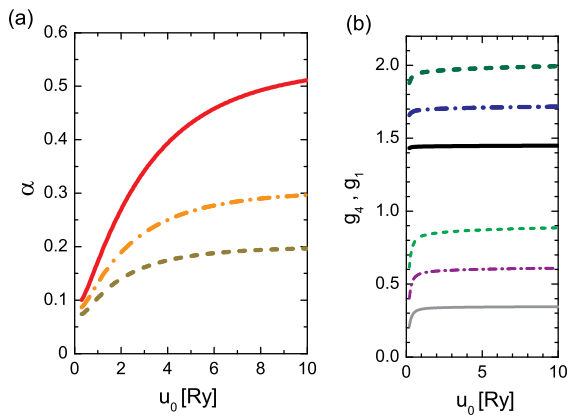}
\caption{(a) 
The TLL exponent $\alpha$ as a function of the field modulation amplitude $u_0$.
The cylinder radius $d$ of the quasi-1D system is taken to be
$d=0.1$ (dashed), $d=0.3$ (dashed-dotted) and $d=1.0$ (solid)
in units of $a_{\rm B}$.
The pamameters $a=5.0$, $b=1.0$, $\kappa = 10^{-4}$ are fixed for all curves.
(b) $u_0$-dependences of $g_4(=g_2)$ (thick curves)
and $g_1$ (thin) under the numerical conditions same as in (a).}
\label{fig_03}
\end{figure}

We demonstrate below that the value of $\alpha$ can be tunable
artificially by imposing an appropriate magnitude of the periodic external field.
Figure \ref{fig_03}(a) shows the field-induced change in the TLL exponent $\alpha$
for different values of the cylinder radius:
$d=0.1$ (dashed), $d=0.3$ (dashed-dotted) and $d=1.0$ (solid)
in units of $a_{\rm B}$.
We fixed the pamameters $a=5.0$, $b=1.0$, 
and $\kappa a_{\rm B} = 10^{-4}$ so as to satisfy the condition
$\kappa \ll k_F$ in accord to the bosonization procedure.\cite{Voit_1995}

The most important observation in Fig.~\ref{fig_03}(a) is 
the monotonic increase in $\alpha$ with increasing $u_0$.
In particular, the data of $\alpha$ for $d=1.0$ exhibits
a pronounced enhancement.
Such the $u_0$-driven shift in $\alpha$ 
is caused by a simple mechanism based on 
the behaviors of constituents $v_F$, $g_4(=g_2)$ and $g_1$
contained in formula (\ref{eq_030}) of ${\cal K}$.
First, it follows from Fig.~\ref{fig_03}(b) that 
$g_4$ and $g_1$ are almost independent of the change in $u_0$
and satisfy the inequality $g_4 > g_1$.
Hence, a decrease in $v_F$ causes a decrease in ${\cal K}$ defined by
Eq.~(\ref{eq_030}).
Second, the above inequality implies ${\cal K} < 1$,
and thus the decrease in ${\cal K}$ results in an increase in $\alpha$ 
as understood from Eq.~(\ref{eq_025}).
As a consequence, $\alpha$ can be raised by enhancing the field modulation amplitude $u_0$.
For $u_0\gg 10.0$, $\alpha$ converges to a limiting value determined by
$\alpha=({\cal K} + {\cal K}^{-1})/2 - 1$ and ${\cal K}=[g_1/(2g_4-g_1)]^{1/2}$,
since $v_F$ vanishes asymptotically.
The upper limit of $\alpha$ is dependent of the geometric parameters
$d$, $a$, $b$;
therefore, it is crucial to set appropriate values of the three parameters
in order to obtain optimal controllability of the exponent $\alpha$.

It should be remarked that we have discussed spinless fermion systems;
they are realized in spin-polarized ultracold fermionic
gases\cite{Gunter_2005}
and in spin-$1/2$ antiferromagnetic ladders,\cite{Chitra_1997,Giamarchi_1999,Hikihara_2001}
whereas many quasi-1D systems involve the effects of the spin degrees of freedom.
When we take into account the spin degrees of freedom,
Eqs.~(\ref{eq_025}) and (\ref{eq_030}) can be rewritten 
because of the presence of SU(2) symmetry
as\cite{Voit_1995}
\begin{eqnarray}
\alpha &=& \frac{1}{4} \left( {\cal K}_\rho + \frac{1}{{\cal K}_\rho} \right) - \frac{1}{2}, \\
{\cal K}_\rho &=&
\left[
\frac
{2 \pi \hbar v_F + 
\left(g_{4 \|} + g_{4 \perp} - g_{2 \|} - g_{2 \perp} + g_{1 \|}\right)}
{2 \pi \hbar v_F + 
\left(g_{4 \|} + g_{4 \perp} + g_{2 \|} + g_{2 \perp} - g_{1 \|}\right)}
\right]^{1/2}.
\end{eqnarray}
Here, $g_{i \|}$ and $g_{i \perp}$ ($i=1,2,4$) express
the matrix elements of interaction between electrons with
parallel spin and those with antiparallel spin, respectively.
In the usual Coulomb interaction, the
matrix elements do not depend on the spin degree of freedom; consequently,
$g_{i \|} = g_{i \perp}$ is expected.
Therefore, we anticipate that the sizeable shift in $\alpha$ demonstrated in
Fig.~\ref{fig_03} should be qualitatively correct even if we take into account the spin degree of freedom.
Details including the actual calculation will be shown elsewhere.

\section{Summary}

We have theoretically shown that the TLL exponent $\alpha$
can be artificially controlled by applying a perodic
external electric field to the system.
The bosonization procedure has been used to obtain
the analytic expression of the exponent $\alpha$
as a function of the amplitude $u_0$ and period $\lambda$ of
the external field modulation $u(z)$.
The result indicates that $\alpha$ increases significantly with
the potential amplitude $u_0$, whose magnitude is within the realm
of the existing experiments.
The significant variation in $\alpha$ is attributed to
the field-induced shift in the Fermi velocity $v_F$ of the single-particle state.
The present results indicate that the quantum transport properties
of quasi-1D systems can be tuned by manipulating $\alpha$.
Experimental confirmation will provide a novel approach to manipulating 1D quantum systems.

\section*{Acknowledgments}

We are grateful to K.~Yakubo, S.~Iwabuchi, J.~Onoe, T.~Ito, and Y.~Toda
for their helpful comments prior to commencement of this study.
HS is thankful for the financial support from the Kazima Foundation
and for the assistance provided by M.~Arroyo at the UPC facility.
This work is supported by a Grant-in-Aid for Scientific
Research from MEXT, Japan,
and Nara Women's University Intramural Grant for Project Research.
Numerical simulations were carried out in part using
the facilities of the Supercomputer Center, ISSP, University of Tokyo.

\appendix

\section{Dispersion relation of 1D periodic systems}

Here, we derive the dispersion relation of Eq.~(\ref{eq_05})
for a 1D system subjected to the stepwise periodic potential given by Eq.~(\ref{eq_02}).
It follows from Eqs.~(\ref{eq_01}) and (\ref{eq_02}) that
$\psi(z)$ behaves as
\begin{equation}
\psi(z) = \left\{
\begin{array}{cc}
A e^{iKz} + B e^{-iKz}, & (0\le z<a) \\ [3pt]
C e^{Qz} + D e^{-Qz}, & (-b \le z < 0)
\end{array}
\right.
\label{eq_03}
\end{equation}
and $\psi(z+a+b)=\psi(z)e^{ik(a+b)}$,
where $\ep = \hbar^2 K^2/(2m^*)$ and $u_0 - \ep = \hbar^2 Q^2/(2m^*)$.
The continuity conditions of $\psi$ and $d\psi/dz$ at both $z=0$ and $z=a$
lead to the matrix equation $\bm{M}\bm{v} = \bm{0}$,
in which $\bm{v} = [A,B,C,D]^T$ and
\begin{equation}
\bm{M} = \left[
\begin{array}{cccc}
1 & 1 & -1 & -1 \\ [3pt]
iK & -iK & -Q & Q \\ [3pt]
\sigma & 1/\sigma & -\omega/\rho & -\rho\omega \\ [3pt]
iK \sigma & -iK/\sigma & -Q\omega/\rho & Q\rho\omega
\end{array}
\right],
\label{eq_06m}
\end{equation}
with $\sigma=e^{iKa}$, $\rho=e^{Qb}$, and $\omega = e^{ik(a+b)}$.
Solving the secular equation ${\rm det}\bm{M} = 0$,
we can obtain the desired result of the dispersion relation (\ref{eq_05}).

In general,
the determinant ${\rm det}\bm{M}$ of a square matrix $\bm{M}$
consisting of $n^2$ elements $\{a_{ij}\}$
is represented by ${\rm det}\bm{M} = \sum_{i=1}^n (-1)^{i+j} a_{ij} C_{ij}$ with an arbitrary fixed value of $j$,
where $C_{ij}$ is the determinant of the $(n-1)\times (n-1)$ matrix
that is obtained from $\bm{M}$ by erasing its $i$th row and $j$th column
(see Ref.~\cite{Higher} for the proof).
Hence, for $Q\ne 0$, the determinant of matrix $\bm{M}$ given by Eq. (\ref{eq_06m}) is
\begin{eqnarray}
& &{\rm det}\bm{M} \nonumber \\ [4pt]
&=&
Qq\omega \left|
\begin{array}{ccc}
1 & 1 & -1 \\
iK & -iK & -Q \\
\sigma & 1/\sigma & -\omega/\rho
\end{array}
\right|
+ \frac{Q\omega}{\rho}
\left|
\begin{array}{ccc}
1 & 1 & -1 \\
iK & -iK & Q \\
\sigma & 1/\sigma & -\rho\omega
\end{array}
\right| \nonumber \\ [4pt]
&-&\frac{iK}{\sigma}
\left|
\begin{array}{ccc}
1 & -1 & -1 \\
iK & -Q & Q \\
\sigma & -\omega/\rho & -\rho\omega
\end{array}
\right|
- iK \sigma
\left|
\begin{array}{ccc}
1 & -1 & -1 \\
-iK & -Q & Q \\
1/\sigma & -\omega/\rho & -\rho\omega
\end{array}
\right| \nonumber \\ [8pt]
&=&
4 iKQ \left( 1 + \omega^2 - 2\omega \cos Ka \cosh Qb \right) \nonumber \\
& &+ 4i \left( K^2 - Q^2\right) \omega \sin Ka \sinh Qb,
\end{eqnarray}
and thus ${\rm det}\bm{M} = 0$ reads as
\begin{equation}
\frac12 \left(\omega + \frac{1}{\omega} \right)
=
\cos Ka \cosh Qb - \frac{K^2 - Q^2}{2KQ} \sin Ka \sinh Qb.
\end{equation}
This result is equivalent to the set of Eqs.~(\ref{eq_05}) and (\ref{eq_06x}).
In addition, the derivative $\pa f/\pa \ep$ is obtained by differentiating
the right side of Eq.~(\ref{eq_06x}) with respect to $\ep$.
Straightforward calculation yields
\begin{eqnarray}
\!\!\!\!\frac{\pa f}{\pa \ep}
&=&\!\!
\frac{\zeta_b}{4 \eta_0} \!\! \left( \frac{\eta_0^2}{\eta_1^2} - \frac{2\zeta_a}{\zeta_b} -1 \right)
\sin (\zeta_a \eta_0) \cosh (\zeta_b \eta_1) \nonumber \\
&+&\!\!
\frac{\zeta_b}{4 \eta_1} \!\! \left( \frac{\eta_1^2}{\eta_0^2} - \frac{2\zeta_b}{\zeta_a} -1 \right)
\cos (\zeta_a \eta_0) \sinh (\zeta_b \eta_1) \nonumber \\
&-&\!\!
\frac{1}{4 \eta_0 \eta_1} \!\! \left( \frac{\eta_1^2}{\eta_0^2} + \frac{\eta_0^2}{\eta_1^2} + 2 \right)
\sin (\zeta_a \eta_0) \sinh (\zeta_b \eta_1).
\label{eq_a012}
\end{eqnarray}

Specifically, when $Q=0$ ({\it i.e.,} $\ep = u_0$),
the continuity conditions of $\psi$ and $d\psi/dz$ at $z=a$
yield $Ka = \ell \pi$ and $k(a+b) = \ell \pi$ with an integer $\ell$.
Given $u_0>0$, therefore, the eigenstate exactly at $\ep = u_0$ exists
only if there is an integer $\ell$ that satisfies
$\hbar^2/(2 m^*)\cdot (\ell\pi/a)^2 = u_0$.
Otherwise, no eigenstate lies at $\ep = u_0$.

Once an eigenenergy $\ep$ and the corresponding $k$ are obtained,
we introduce them into the equation $\bm{M}\bm{v}=0$ to evaluate
the coefficients $A,B,C,D$ of the eigenfunction $\psi(z)$
that belongs to $\ep$.
The explicit forms of the coefficients are
\begin{eqnarray}
\frac{B}{A}
&=&
- \frac{Q+iK}{Q-iK} \cdot \frac{e^{ik(a+b)} e^{-Qb} - e^{iKa}}{e^{ik(a+b)} e^{-Qb} - e^{-iKa}}, \\ [5pt]
\frac{C}{A}
&=&
\frac{Q+iK}{Q} \cdot \frac{i\sin Ka}{e^{ik(a+b)} e^{-Qb} - e^{-iKa}},
\end{eqnarray}
and $D/A = 1+(B/A)-(C/A)$; $A$ is uniquely determined by the normalization
condition of $\psi(z)$.

\section{Proof of Eq.~(\ref{eq_030})}

The TLL parameter ${\cal K}$ introduced in Eq.~(\ref{eq_030}) is deduced from
the momentum representation of the two-electron interaction Hamiltonian $\hat{{\cal H}}_{\rm int}$
for a quasi-1D periodic system.
In the real-space representation, $\hat{{\cal H}}_{\rm int}$ is given by
\begin{eqnarray}
\hat{{\cal H}}_{\rm int}
&=&
\frac12 \int d\bm{r}_1 \int d\bm{r}_2
V\left( \left| \bm{r}_1 - \bm{r}_2 \right| \right) \nonumber \\
& & \qquad
\times
\hat{\Psi}^{\dag}(\bm{r}_1)
\hat{\Psi}^{\dag}(\bm{r}_2)
\hat{\Psi}(\bm{r}_2)
\hat{\Psi}(\bm{r}_1),
\label{eq_appb1}
\end{eqnarray}
where
\begin{equation}
\hat{\Psi}(\bm{r}_i) = \sum_{k} \hat{c}_{k} \psi_k(z_i) \chi \left( \bm{r}_{\perp}^i \right)
\label{eq_appb2}
\end{equation}
with a fermionic annihilation operator $\hat{c}_k$.
Substituting Eq.~(\ref{eq_appb2}) into Eq.~(\ref{eq_appb1}),
we have
\begin{equation}
\hat{{\cal H}}_{\rm int} = \frac12
\sum_{k_1,k_2,k_3,k_4} \tilde{V}_{k_1,k_2;k_3,k_4}
\hat{c}_{k_1}^{\dag} \hat{c}_{k_2}^{\dag} \hat{c}_{k_3} \hat{c}_{k_4},
\label{eq_appb15}
\end{equation}
and
\begin{eqnarray}
\tilde{V}_{k_1,k_2;k_3,k_4}
&=&
\int \!{\cal D}\bm{r}_{\perp} \int\! dz_1 \int\! dz_2
V\left(|\bm{r}_1 \!-\! \bm{r}_2|\right) \nonumber \\
&\times&
\psi^*_{k_1}(z_1) \psi^*_{k_2}(z_2) \psi_{k_3}(z_2) \psi_{k_4}(z_1),
\label{eq_042}
\end{eqnarray}
in which the integration operator
$\int {\cal D}\bm{r}_{\perp} \equiv \int \!d\bm{r}_{\perp}^1 \int \!d\bm{r}_{\perp}^2
\left| \chi\left(\bm{r}_{\perp}^1 \right) \right|^2
\left| \chi\left(\bm{r}_{\perp}^2 \right) \right|^2$
acts on $V\left(|\bm{r}_1 \!-\! \bm{r}_2|\right)$
since $\bm{r}_i$ depends on $\bm{r}_{\perp}^i$.
In the meantime, we derive an alternative form of $\tilde{V}_{k_1,k_2;k_3,k_4}$
({\it i.e.,} Eq.~(\ref{eq_appb12}))
which allows to obtain 
the momentum representation of $\hat{\cal{H}}_{\rm int}$
({\it i.e.,} Eq.~(\ref{eq_appb5}))
relevant to the formula of ${\cal K}$.

Bearing in mind the periodicity of a system with period $\lambda = a+b$,
we replace the variables $z_1, z_2$ ($0\le z_i \le L$, $i=1,2$) in Eq.~(\ref{eq_042})
with $z_1+l'\lambda$ and $z_2+(l'-l) \lambda$ ($0<z_i<\lambda$, $i=1,2$),
respectively;
the integers $l$ and $l'$ are independent of each other
and vary from 0 to $N-1$.
Applying the periodic boundary condition of $\psi_k(z)=\psi_k(z+L)$ for an arbitrary $k$,
we obtain
\begin{eqnarray}
\tilde{V}_{k_1,k_2;k_3,k_4}\!\!
&=&
\!\!\frac{1}{N^2} \sum_l \sum_{l'} \int {\cal D}\bm{r}_{\perp}
\int_U \!dz_1 \int_U \!dz_2 \nonumber \\
&\times& \!\! V\left(|\bm{r}_1 \!-\! \bm{r}_2 \!+\! l\lambda \bm{e}_z|\right) \bar{\phi}(z_1,z_2) \nonumber \\
&\times& e^{-i(k_1-k_4)z_1} e^{-i(k_2-k_3)z_2} \nonumber \\
&\times& e^{-i(k_1 + k_2 - k_3 - k_4)l'\lambda} e^{i(k_2-k_3)l\lambda},
\label{eq_appb8}
\end{eqnarray}
where
$\bar{\phi}(z_1,z_2) \equiv \phi_{k_1}^*(z_1) \phi_{k_2}^*(z_2) \phi_{k_3}(z_2) \phi_{k_4}(z_1)$
and $\int_U$ symbolizes the integration within the unit cell $z = [0, \lambda]$;
$\bm{e}_z$ is a unit vector in the axial direction.
Summation over $l'$ in Eq.~(\ref{eq_appb8}) yields a term $N \delta_{k_1+k_2, k_3+k_4}$
(we consider only normal scattering contributions, omitting the Umklapp scattering
processes).
We further assume a momentum conservation that requires relabeling
of the set $(k_1, k_2, k_3, k_4)$ by $(k_1+q, k_2-q, k_2, k_1)$.
Introducing the assumption into Eq.~(\ref{eq_appb8})
and performing summation with respect to $l$,
we find that $\tilde{V}(k_1, k_2; \; q) \equiv \tilde{V}_{k_1+q,k_2-q;k_2,k_1}$ is given by
\begin{eqnarray}
\!\!\!\!\!\!\!\!\!\!
& &\tilde{V}(k_1, k_2; \; q)
=
\frac1N \! \int \! {\cal D}\bm{r}_{\perp} \int \! dz_1 \int_U \! dz_2
V\left(|\bm{r}_1 \!-\! \bm{r}_2|\right) \nonumber \\
\!\!\!\!\!\!\! & & \;\; \times
\phi_{k_1+q}^*(z_1) \phi_{k_1}(z_1)e^{-iqz_1}
\phi_{k_2-q}^*(z_2) \phi_{k_2}(z_2)e^{iqz_2}. \label{eq_appb12} \\
& & \nonumber
\end{eqnarray}
In Eq.~(\ref{eq_appb12}),
the subscript $U$ originally attached to $\int_U dz_1$ vanishes
so that $z_1$ ranges from $0$ to $L$.

We are ready to evaluate the TLL parameter ${\cal K}$.
From Eq.~(\ref{eq_appb12}) and Eq.~(\ref{eq_appb15}),
it follows that
\begin{equation}
\hat{{\cal H}}_{\rm int} = \frac12 \!\! \sum_{k_1, k_2, q} \!\!\!
\tilde{V}(k_1, k_2; \; q)
\hat{c}_{k_1+q}^{\dag} \hat{c}_{k_2-q}^{\dag} \hat{c}_{k_2} \hat{c}_{k_1}.
\label{eq_app075}
\end{equation}
Substituting $k_i = p_i k_F + \kappa$ into Eq.~(\ref{eq_app075})
under the assumption that of $|\kappa| \ll k_F$ and $p=\pm 1$, we get
\begin{eqnarray}
\hat{{\cal H}}_{\rm int} &=& \frac12 \sum_p \sum_{\kappa,\kappa',q} \left[
\tilde{V}(pk_F, pk_F; \; 0) \hat{c}_{p, \kappa+q}^{\dag} \hat{c}_{p, \kappa'-q}^{\dag} \hat{c}_{p,\kappa'} \hat{c}_{p,\kappa} \right. \nonumber \\
&+&
\tilde{V}(pk_F, -pk_F; \; -2pk_F) \hat{c}_{p, \kappa+q}^{\dag} \hat{c}_{-p, \kappa'-q}^{\dag} \hat{c}_{p,\kappa'} \hat{c}_{-p,\kappa} \nonumber \\
&+&
\left. \tilde{V}(pk_F, -pk_F; \; 0) \hat{c}_{p, \kappa+q}^{\dag} \hat{c}_{-p, \kappa'-q}^{\dag} \hat{c}_{-p,\kappa'} \hat{c}_{p,\kappa} \right],
\end{eqnarray}
where $\hat{c}_{p,\kappa} = \hat{c}_{p k_F + \kappa}$.
Since the time reversal symmetry of the system assures that
$\tilde{V}_{k_1,k_2; k_3,k_4} = \tilde{V}_{-k_1,-k_2; -k_3,-k_4}$,
we have $\tilde{V}(k_1, k_2; \; q) = \tilde{V}(-k_1-q, -k_2+q; \; q)$
that implies
\begin{eqnarray}
\tilde{V}(k_F, k_F; \; 0) &=& \tilde{V}(-k_F, -k_F; \; 0), \\
\tilde{V}(k_F, -k_F; \; 0) &=& \tilde{V}(-k_F, k_F; \; 0).
\end{eqnarray}
As a consequence, we obtain the result
\begin{eqnarray}
\!\!\!\!\!\!\!\!\!\!\! \hat{{\cal H}}_{\rm int}
&=&
\frac{g_4}{2L} \sum_p \sum_{\kappa,\kappa',q}
\hat{c}_{p, \kappa+q}^{\dag} \hat{c}_{p, \kappa'-q}^{\dag} \hat{c}_{p,\kappa'} \hat{c}_{p,\kappa} \nonumber \\
&+&
\frac{g_2-g_1}{2L} \sum_p \sum_{\kappa,\kappa',q}
\hat{c}_{p, \kappa+q}^{\dag} \hat{c}_{-p, \kappa'-q}^{\dag} \hat{c}_{-p,\kappa'} \hat{c}_{p,\kappa},
\label{eq_appb5}
\end{eqnarray}
where $g_4$, $g_2$, and $g_1$ are defined by Eqs.(\ref{eq_g4})-(\ref{eq_g1}).
Through the diagonalization of $\hat{{\cal H}}_{\rm int}$ in Eq.~(\ref{eq_appb5}),
we finally obtain the TLL parameter ${\cal K}$ given in Eq.~(\ref{eq_030});
this parameter depends on 
the two coefficients $g_4$ and $g_2-g_1$ prior to the summations
in Eq.~(\ref{eq_appb5}).

%
%


\begin{thebibliography}{99}

\bibitem{Tomonaga}
S.~Tomonaga, Prog.~Theor.Phys. {\bf 5}, 544 (1950).
\bibitem{Luttinger}
J.~M.~Luttinger, Phys.~Rev. {\bf 119}, 1153 (1960);
J.~Math.~Phys. {\bf 4}, 1154 (1963).
\bibitem{Voit_1995}
J.~Voit, Rep.~Prog.~Phys. {\bf 58}, 977 (1995).

\bibitem{Bockrath_1999}
M.~Bockrath, D.~H.~Cobden, J.~Lu, A.~G.~Rinzler, R.~E.~Smalley, L.~Balents,
and P.~L.~McEuen, Nature (London) {\bf 397}, 598 (1999).
\bibitem{Yao_1999}
Z.~Yao, H.~Postma, L.~Balents, and C.~Dekker, Nature (London) {\bf 402}, 273 (1999).
\bibitem{Bachtold_2001}
A.~Bachtold, M.~de Jonge, K.~Grove-Rasmussen, P.~L.~McEuen, M.~Buitelaar, and C.~Schonenberger,
Phys.~Rev.~Lett. {\bf 87}, 166801 (2001).
\bibitem{Ishii_2003}
H.~Ishii, H.~Kataura, H.~Shiozawa, H.~Yoshioka, H.~Otsubo, Y.~Takayama, T.~Miyahara, S.~Suzuki, Y.~Achiba, M.~Nakatake, T.~Narimura, M.~Higashiguchi, K.~Shimada, H.~Namatame, and M.~Taniguchi, Nature (London) {\bf 426}, 540 (2003);
H.~Yoshioka, Physica E {\bf 18}, 212 (2003).
\bibitem{Tombros_2006}
N.~Tombros, S.~J.~van der Molen, and B.~J.~van Wees, Phys.~Rev.~B, {\bf 73}, 233403 (2006).
\bibitem{Auslaender_2002}
O.~M.~Auslaender, A.~Yacoby, R.~de Picciotto, K.~W.~Baldwin, L.~N.~Pfeiffer, K.~W.~West,
Science {\bf 295}, 825 (2002).
\bibitem{Tserkovnyak_2002_2003}
Y.~Tserkovnyak, B.~I.~Halperin, O.~M.~Auslaender and A.~Yacoby,
Phys.~Rev.~Lett. {\bf 89}, 136805 (2002); Phys.~Rev.~B {\bf 68}, 125312 (2003).
\bibitem{Auslaender_2003}
O.~M.~Auslaender, H.~Steinberg, A.~Yacoby, Y.~Tserkovnyak, B.~I.~Halperin, K.~W.~Baldwin,
L.~N.~Pfeiffer and K.~W.~West, Science {\bf 308}, 88 (2005).
\bibitem{Steinberg_2008}
H.~Steinberg, G.~Barak, A.~Yacoby, L.~N.~Pheiffer, K.~W.~West, B.~I.~Halperin, and K.~L.~Hur, Nature Phys. {\bf 4}, 116 (2008).
\bibitem{Jompol_2009}
Y.~Jompol, C.~J.~B.~Ford, J.~P.~Griffiths, I.~Farrer, G.~A.~C.~Jones, D.~Anderson, D.~A.~Ritchie, T.~W.~Silk, and A.~J.~Schofield,
Science {\bf 325}, 597 (2009).
\bibitem{Schwartz_1998}
A.~Schwartz, M.~Dressel, G.~Gr\"uner, V.~Vescoli, L.~Degiorgi, and T~ Giamarchi,
Phys.~Rev.~B {\bf 58}, 1261 (1998).
\bibitem{Claessen_2002}
R.~Claessen, M.~Sing, U.~Schwingenschl\"ogl, P.~Blaha, M.~Dressel,
and C.~S.~Jacobsen, Phys.~Rev.~Lett. {\bf 88}, 096402 (2002).
\bibitem{Sing_2003} M.~Sing, U.~Schwingenschl\"ogl, R.~Claessen, P.~Blaha,
J.~M.~P.~Carmelo, L.~M.~Martelo, P.~D.~Sacramento, M.~Dressel, and C.~S.~Jacobsen,
Phys.~Rev.~B, {\bf 68}, 125111 (2003).
\bibitem{Chang_1996}
A.~M.~Chang, L.~N.~Pfeiffer, and K.~W.~West, Phys.~Rev.~Lett. {\bf 77}, 2538 (1996).
\bibitem{CKim_1996}
C.~Kim, A.~Y.~Matsuura, Z.~X.~Shen, N.~Motoyama, H.~Eisaki, S.~Uchida, T.~Tohyama, and S.~Maekawa,
Phys.~Rev.~Lett. {\bf 77}, 4054 (1996).
\bibitem{Segovia_1999}
P.~Segovia, D.~Purdie, M.~Hengsberger, and Y.~Baer, Nature (London) {\bf 402}, 504 (1999).
\bibitem{Slot_2004}
E.~Slot, M.~A.~Holst, H.~S.~J.~van der Zant, and S.~V.~Zaitsev-Zotov,
Phys.~Rev.~Lett. {\bf 93}, 176602 (2004).
\bibitem{Aleshin_2004}
A.~N.~Aleshin, H.~J.~Lee, Y.~W.~Park, and K.~Akagi, Phys.~Rev.~Lett. {\bf 93}, 196601 (2004).
\bibitem{Hager_2005}
J.~Hager, R.~Matzdorf, J.~He, R.~Jin, D.~Mandrus, M.~A.~Cazalilla, and E.~W.~Plummer,
Phys.~Rev.~Lett. {\bf 95}, 186402 (2005).
\bibitem{Venkataraman_2006}
L.~Venkataraman, Y.~S.~Hong, and P.~Kim, Phys.~Rev.~Lett. {\bf 96}, 076601 (2006).
\bibitem{BJKim_2006}
B.~J.~Kim, H.~Koh, E.~Rotenberg, S.~J.~Oh, H.~Eisaki, N.~Motoyama, S.~Uchida, T.~Tohyama, S.~Maekawa, Z.~X.~Shen and C.~Kim,
Nature Phys. {\bf 2}, 397 (2006).

\bibitem{Yuen_2009}
J.~D.~Yuen, R.~Menon, N.~E.~Coates, E.~B.~Namdas, S.~Cho, S.~T.~Hannahs, D.~Moses,
and A.~J.~Heeger, Nature Mater. {\bf 8}, 572 (2009).
\bibitem{Imambekov_2009}
A.~Imambekov, L.~I.~Glazman, Science {\bf 323}, 228 (2009);
Phys.~Rev.~Lett. {\bf 102}, 126405 (2009).
\bibitem{Tokuno_2008}
A.~Tokuno, M.~Oshikawa, and E.~Demler, Phys.~Rev.~Lett. {\bf 100}, 140402 (2008).
\bibitem{Bubanja_2009}
V.~Bubanja and S.~Iwabuchi, Phys.~Rev.~B {\bf 79}, 035312 (2009).
\bibitem{Shima_2009}
H.~Shima, H.~Yoshioka and J.~Onoe,
Phys.~Rev.~B {\bf 79}, 201401(R) (2009);
Physica E (2010) {\it in press}. [arXiv:0908.2565]

\bibitem{Fabrizio_1995}
M.~Fabrizio and A.~O.~Gogolin, Phys.~Rev.~B {\bf 51}, 17827 (1995).

\bibitem{Postma_2000}
H.~W.~C.~Postma, M.~de Jonge, Z.~Yao, and C.~Dekker,
Phys.~Rev.~B {\bf 62}, R10653 (2000).

\bibitem{Bellucci_2006}
S.~Bellucci, J.~Gonz\'alez, P.~Onorato, and E.~Perfetto, Phys.~Rev.~B {\bf 74}, 045427 (2006).

\bibitem{Klanjsek_2008}
M.~Klanj\v{s}ek, H.~Mayaffre, C.~Berthier, M.~Horvati\'c, B.~Chiari, O.~Piovesana,
P.~Bouillot, C.~Kollath, E.~Orignac, R.~Citro, and T.~Giamarchi,
Phys.~Rev.~Lett. {\bf 101}, 137207 (2008).

\bibitem{Ono}
S.~Ono and H.~Shima, Phys.~Rev.~B {\bf 79}, 235407 (2009);
Physica E (2010) {\it in press}. [arXiv:arXiv:0909.1135]

\bibitem{Taira}
H.~Taira and H.~Shima, J.~Phys.:~Condens.~Mat. (2010) {\it in press}.
[arXiv:0904.3149]

\bibitem{Adachi_1985}
S.~Adachi, J.~Appl.~Phys. {\bf 58}, R1 (1985).


\bibitem{Tanatar_1989}
B.~Tanatar and D.~M.~Ceperley, Phys.~Rev.~B {\bf 39}, 5005 (1989).
\bibitem{Shulenburger_2008}
L.~Shulenburger, M.~Casula, G.~Senatore and R.~M.~Martin, Phys.~Rev.~B {\bf 78}, 165303 (2008).
\bibitem{Gunter_2005}
K.~G\"unter, T.~St\"oferle, H.~Moritz, M.~K\"ohl, and T.~Esslinger,
Phys.~Rev.~Lett. {\bf 95}, 230401 (2005).
\bibitem{Chitra_1997}
R.~Chitra and T.~Giamarchi, Phys.~Rev.~B {\bf 55}, 5816 (1997).
\bibitem{Giamarchi_1999}
T.~Giamarchi and A.~M.~Tsvelik, Phys.~Rev.~B {\bf 59}, 11398 (1999).
\bibitem{Hikihara_2001}
T.~Hikihara and A.~Furusaki, Phys.~Rev.~B {\bf 63}, 134438 (2001).
\bibitem{Higher} H.~Shima and T.~Nakayama, {\it Higher Mathematics for Physics and Engineering},
(2010) (Springer-Verlag).
\end{thebibliography}
\end{document}